\documentclass[prl,twocolumn,showkeys,showpacs,superscriptaddress,amsmath,amsfonts]{revtex4-1}
\usepackage[utf8x]{inputenc}
\usepackage{microtype}
\usepackage{lmodern}
\usepackage{graphicx}
\usepackage{hyperref}% add hypertext capabilities
\hypersetup{
    colorlinks=true,
    linkcolor=blue,
    filecolor=magenta,      
    urlcolor=red,
    citecolor=blue,
}
\usepackage{color}
\usepackage{todonotes}
\usepackage{bm}
\usepackage{nicefrac}
\usepackage{braket}

\newcommand{\mean}[1]{\langle#1\rangle}
\newcommand{\vect}[1]{\bm{{#1}}}
\newcommand{\ssum}{{\Sigma}}
\newcommand{\pdag}{{\phantom\dag}}
\begin{document}

\title{Theory of dispersive optical phonons in resonant inelastic x-ray scattering experiments}

\author{Krzysztof Bieniasz}
\email{krzysztof.t.bieniasz@gmail.com}
\affiliation{\mbox{Quantum Matter Institute, University of British Columbia,
             Vancouver, British Columbia, Canada V6T 1Z4}}
\affiliation{Department of Physics and Astronomy, University of British Columbia, Vancouver, British Columbia, Canada V6T 1Z1}

\author{Steve Johnston}
\email{sjohn145@utk.edu}
\affiliation{Department of Physics and Astronomy, University of Tennessee, Knoxville, Tennessee 37996, USA}
\affiliation{Institute for Advanced Materials and Manufacturing, University of Tennessee, Knoxville, Tennessee 37996, USA\looseness=-1}

\author{Mona Berciu}
\email{berciu@phas.ubc.ca }
\affiliation{\mbox{Quantum Matter Institute, University of British Columbia, Vancouver, British Columbia, Canada V6T 1Z4}}
\affiliation{Department of Physics and Astronomy, University of British Columbia, Vancouver, British Columbia, Canada V6T 1Z1}

\date{\today}

\begin{abstract}
The community currently lacks a complete understanding of how resonant inelastic x-ray scattering (RIXS) experiments probe the electron-phonon ($e$-ph) interaction in solids. For example, most theoretical models of this process have focused on dispersionless Einstein phonons. Using a recently developed momentum average (MA) variational approximation for computing RIXS spectra of band insulators, we examine the influence of both electron and phonon dispersion in the intermediate state of the scattering process. 
We find that the inclusion of either, and their mutual interplay, introduces significant momentum variations in the RIXS intensity, even for momentum-independent electron-phonon coupling. The phonon dispersion also induces nontrivial changes in the excitation line shapes, which can have a quantitative impact on the data analysis. These results highlight the considerable challenges of interpreting RIXS data in actual materials.
\end{abstract}

\maketitle

\noindent\textit{Introduction} --- 
Resonant inelastic x-ray scattering (RIXS) \cite{KotaniRMP2001, AmentRMP2011} is being used increasingly to study electron-phonon ($e$-ph) coupling 
in solids. This application is being driven by the steady improvements of both the instrument resolution and our understanding of the RIXS cross-section. For example, theoretical modeling has suggested that RIXS can access the $e$-ph coupling strength with momentum resolution and element specificity~\cite{AmentEPL2011, LeePRL2013, DevereauxPRX2016}.

One of the most popular methods for quantitatively analyzing lattice excitations in RIXS spectra is the single-site framework developed by Ament \emph{et al.}~\cite{AmentEPL2011}. It approximates the infinite system with a single isolated site whose local electron density in the valence orbital couples to the lattice displacements. This simplified model's exact RIXS scattering amplitude can be computed within the Kramers-Heisenberg formalism using a Lang-Firsov transformation. Its key predictions are that the $e$-ph coupling produces a series of low-energy harmonic excitations in the energy loss spectra, whose relative intensities can be mapped onto the strength of the $e$-ph interaction. 

While this single-site model has been widely employed for data analysis~\cite{MeyersPRL2018, RossiPRL2019, BraicovichPRR2020, PengPRL2020, peng2021dopingdependence}, its approximations are drastic, and it is unclear how relaxing them may affect 
the results of the analysis. For this reason, several groups have attempted to develop alternative approaches. Examples (in no particular order) include generalizations of the single-site framework to include multiple modes or changes in the harmonic potential in the intermediate state~\cite{GeondzhianPRB2020}, exact diagonalization of small clusters~\cite{LeePRL2013, JohnstonNatComm2016}, diagrammatic approaches~ \cite{DevereauxPRX2016, HuangPRX2021}, cumulant expansions of the Green's function~\cite{dashwood2021probing}, and dynamical mean-field theory \cite{WernerPRB2021}.

Recently, we introduced an efficient variational method for computing RIXS spectra for band insulators \cite{BieniaszSciPost2021}. Our method is built on the Momentum Average (MA) class of variational approximations \cite{BerciuPRL2006, BerciuPRB2007} and allows us to treat situations where a core electron is excited into an empty band in the intermediate state of the RIXS process and is allowed to interact with the lattice. We showed that the single-site approximation becomes inaccurate for shallower core-hole potentials and found that the itinerancy of the valence electron leads to momentum dependence in the intensity of the RIXS phonon peaks even if both the $e$-ph coupling and the phonon's dispersion are momentum-independent. Naturally, this raises the question of whether the momentum-dependence of various features in the RIXS spectra can be used to infer momentum dependence of either the $e$-ph coupling and/or phonon dispersion.

Here, we answer the latter part of this question. We extend the MA formalism to study band insulators where the excited electron couples to \textit{dispersive} optical phonons. Our goal is to understand how a finite phonon bandwidth affects the RIXS spectra when the $e$-ph coupling is momentum independent (Holstein). We find that the phonon bandwidth produces specific $\vect{q}$-dependence of the multi-phonon excitations so that 
even for a Holstein coupling, we obtain single- and multi-phonon excitations whose peak location and intensity vary significantly around the first Brillouin zone (BZ). Moreover, the predicted multi-phonon line shapes are complicated, deviating considerably from the Lorentzian or Gaussian shapes frequently adopted when fitting experimental data. These results expand our knowledge of how the details of the $e$-ph coupling are encoded in the RIXS cross-section and further underscore the need to move beyond single-site models in data analysis.\\

\noindent\textit{The Model} --- We examine the RIXS spectra for a band insulator whose valence electrons are coupled to a dispersive optical phonon branch in the intermediate state of the scattering process. The Hamiltonian  is $\mathcal{H} = \mathcal{H}_\mathrm{e} + 
\mathcal{H}_\mathrm{ph} + \mathcal{H}_{\textrm{$e$-ph}} + \mathcal{H}_\mathrm{ch}$. Here,    $\mathcal{H}_{e} = -t\sum_{\mean{ij}}[d_{i}^{\dag}d_{j}^{\pdag}+\mathrm{H.c.}]
  = \sum_{\vect{k}} \epsilon^{\phantom\dagger}_{\vect{k}}d_{\vect{k}}^{\dag}d_{\vect{k}}^{\pdag}$ where $d_{i}^{\dagger}$ ($d_{i}^{\pdag}$) creates (annihilates) an electron at site $i$ in the valence band,
 and  $\epsilon_{\vect{k}}$ is the valence band dispersion.  The optical phonon is described by
  $ \mathcal{H}_\mathrm{ph} = \sum_{\vect{q}} \omega^{\phantom\dagger}_{\vect{q}} b_{\vect{q}}^{\dag}b_{\vect{q}}^{\pdag}$, where $b_{\vect{q}}^{\dag}$ ($b_{\vect{q}}^{\pdag}$) creates (annihilates) a phonon with energy $\omega_{\vect{q}}$ (we set $\hbar=1$). Throughout, we assume that the system is a two-dimensional (2D) square lattice with $a=1$ so that $\epsilon_{\vect{k}} = -2t\left(\cos k_x + \cos k_y\right)$ while $\omega_{\vect{q}} = \omega_{0} + 2\omega_{1}\left(\cos q_{x}+\cos q_{y}\right)$. Since we have in mind optical oxygen modes in transition metal oxides, we further assume that the phonon bandwidth is small, $\omega_{1}\ll\omega_{0}$. 
  The Holstein $e$-ph coupling is 
\begin{equation}
 % \label{eq:mod}
 % \mathcal{H}_{e} &= -t\sum_{\mean{ij}}[d_{i}^{\dag}d_{j}^{\pdag}+\mathrm{H.c.}]
 % = \sum_{\vect{k}} \epsilon^{\phantom\dagger}_{\vect{k}}d_{\vect{k}}^{\dag}d_{\vect{k}}^{\pdag},\\
  %\mathcal{H}_\mathrm{ph} &= \sum_{\vect{q}} \omega^{\phantom\dagger}_{\vect{q}} b_{\vect{q}}^{\dag}b_{\vect{q}}^{\pdag}
  \label{eq:epcp}
  \mathcal{H}_{\textrm{$e$-ph}} = \frac{g}{\sqrt{N}} \sum_{i} e^{-\mathrm{i}\vect{k}\cdot\vect{R}_{i}} d_{i}^{\dag}d_{i}^{\pdag}(b_{\vect{k}}^{\dag}+b_{-\vect{k}}^{\pdag}),
\end{equation}
where $g$ is the strength of the coupling and $N$ is the number of lattice sites. Finally
\begin{equation}
  \label{eq:U}
  \mathcal{H}_\mathrm{ch} = \epsilon_\mathrm{ch}\sum_{i}p_i^\dag p^{\pdag}_i -U_Q\sum_{i}d_{i}^{\dag}d_{i}^{\pdag}\left(1-p_i^\dag p^{\pdag}_i\right)
\end{equation}
describes the core-hole and its interaction with the valence electron. Specifically,  $p_{i}^{\dagger}$ ($p_{i}^{\pdag}$) creates (annihilates) an electron in the relevant core level at site $i$,  $\epsilon_\mathrm{ch}$ is the on-site energy of the core level, and $-U_Q$ is the local \emph{attractive} interaction between the valence electron and the core hole. 
We use a mixed notation for the Holstein coupling [Eq.~(\ref{eq:epcp})], where the electron (phonon) operators are represented in real (momentum) space, for later convenience. Finally, Eq.~(\ref{eq:U}) captures the core-hole's influence on the system in the intermediate state via a local core-hole potential. We note that although the bare $U_Q$ potential is purely local, the effective potential can extend over the entire lattice (albeit decaying very fast away from the core hole)~\cite{Ber10, Ebr12, BieniaszSciPost2021} once dressed by the $e$-ph interaction. \\

\noindent\textit{The Method} --- Our starting point is the standard  Kramers-Heisenberg (KH) equation for the RIXS intensity~\cite{KotaniRMP2001, AmentRMP2011}, which we reformulate by expanding the delta function as the imaginary part of a final state Green's function~\cite{Nocera2018}
\begin{equation}
  \label{eq:kram}
  I(\omega,\vect{q}) = -\frac{1}{\pi}\Im
  \sum_{f} \frac{\lvert F_{fg}({\vect q}, z)\rvert^{2}}{\omega+\mathrm{i}\eta-E_{f}+E_{g}}.  
\end{equation}
Here, $\eta$ is a broadening parameter and $F_{fg}$ is the scattering amplitude   
\begin{equation}
    F_{fg}(\vect{q},z) = 
    \sum_{n,i} e^{\mathrm{i}{\vect q}\cdot {\vect R}_i} 
    \frac{\bra{f}D_i^\dag\ket{n}\bra{n}D_i^\pdag\ket{g}}
    {E_g - E_n + z},
\end{equation}
where $\ket{g}$, $\ket{n}$, and $\ket{f}$ are the initial, intermediate, and final states of the RIXS process with energies $E_g$, $E_n$, and $E_f$, respectively, $z = \omega_\mathrm{in} + \mathrm{i}\Gamma$, $\omega_\mathrm{in}$ and $\omega_\mathrm{out}$ are in energies of the incident and scattered x-ray, $\omega = \omega_\mathrm{out}-\omega_\mathrm{in}$ and ${\vect q}$ are the energy and momentum transferred to the sample, $\Gamma$ is the inverse core-hole lifetime, and $D_i$ is the dipole operator. (Here, we have omitted the geometric prefactors associated with the dipole matrix elements to focus on the effects of the electron and phonon dispersions.)
%are unimportant for our discussion and have been omitted for brevity. 
The specific elemental edge does not matter at our level of modeling~\cite{BieniaszSciPost2021}. 

We briefly explain here our variational approach, delegating all details to the Supplementary Material~\cite{supplement}. The first step is to evaluate the spectral amplitude $F_{fg}$. Following Ref.~\citenum{BieniaszSciPost2021}, we cast it as a generalized propagator  
\begin{equation}
  \label{eq:spec}
  \mathcal{F}_{fg}(\vect{q},z) = \frac{1}{\sqrt{N}}\sum_{i}e^{\mathrm{i}\vect{q}\cdot{\vect R}_{i}} \bra{f}p_{i}^{\dag}d_{i}^{\pdag} \mathcal{G}(z) d_{i}^{\dag}p_{i}\ket{g},
\end{equation}
where $\mathcal{G}(z)=[z-\mathcal{H}+E_g]^{-1}$. 
We generate an equation of motion (EOM) for $\mathcal{F}_{fg}(z)$ by applying the Dyson identity $\mathcal{G}(z) = \mathcal{G}_{0}(z)+\mathcal{G}(z)\mathcal{V}\mathcal{G}_{0}(z)$,
where $\mathcal{V} = \mathcal{H}_{e-\mathrm{ph}}$ and $\mathcal{H}_0= \mathcal{H}-\mathcal{H}_{e-\mathrm{ph}} $ has the associated resolvent $\mathcal{G}_{0}(\omega)$. The EOM for $\mathcal{F}_{fg}(z)$ depends on new propagators, whose EOMs depend on other new propagators, {\it etc.}, generating an infinite hierarchy of coupled EOMs. To simplify it and then solve it,  we define a variational Hilbert space characterized by the size and spread of the phonon cloud~\cite{BerciuPRB2007}, and only keep in the hierarchy the EOMs for propagators consistent with this variational choice. In particular, it has been well documented that for a Holstein coupling that is not deep into the adiabatic regime, a one-site cloud approximation is very accurate (deep in the adiabatic regime, Holstein polaron clouds spread over several consecutive sites and the variational space needs to be expanded accordingly \cite{MarchandPRB2017, CarbonePRB2021}. We implement this one-site cloud variational solution here. We emphasize that this polaron cloud can appear anywhere in the system, it is not restricted to the core-hole site. Additional technical details can be found in the Supplementary Material~\cite{supplement}, as well as Refs.~\citenum{BerciuPRL2006,BerciuPRB2007,BieniaszSciPost2021}.\\

\begin{figure*}[ht!]
  \includegraphics[width=\textwidth]{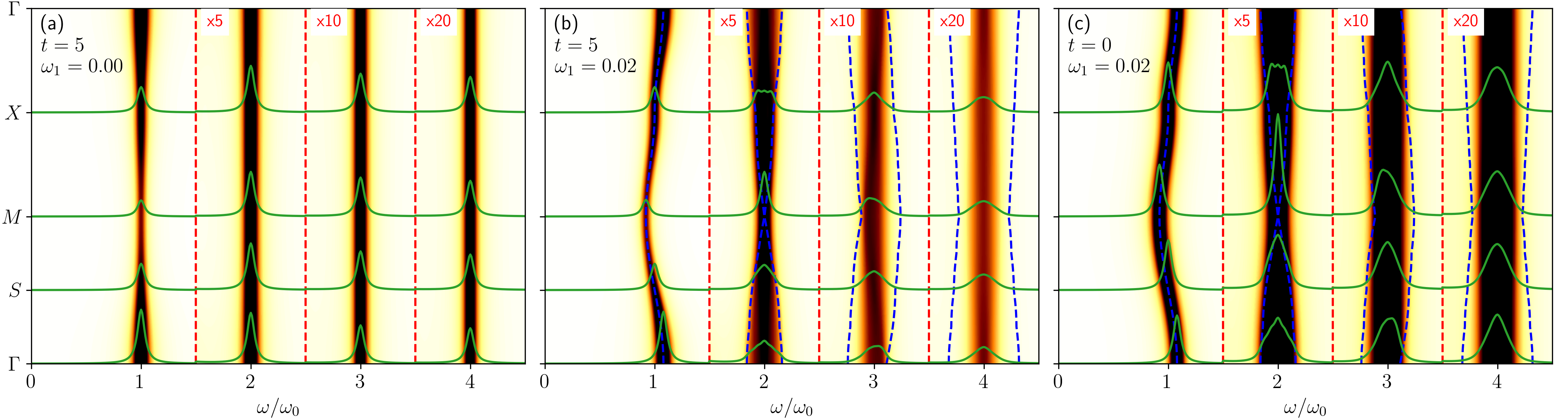}
  \caption{\label{fig:1}RIXS spectra calculated for various model parameters. The results 
  plotted for momentum transfers $\vect{q}$ along the high-symmetry cuts of the first
  Brillouin zone of the square lattice. The high-symmetry points are 
  denoted as $\Gamma=(0,0)$, $M=(1,1)$, $X=(1,0)$, and $S = (1/2,1/2)$ 
  in units of $\pi/a$. All results were obtained with a variational constraint $p = 2$ (see the Supplementary Material~\cite{supplement} for more details). 
  (a) mobile electron coupled to an Einstein phonon ($t=5$, $\omega_{1}=0$); (b) mobile electron 
  coupled to a dispersive optical phonon ($t = 5$, $\omega_1 = 0.02)$;
  (c) localized electron coupled to a dispersive optical phonon branch  ($t = 0$, $\omega_1 = 0.02$).  
  All parameters are in units of $\omega_0=1$. 
  We've multiplied the intensity of the multi-phonon peaks by the factor indicated in the corner of the respective region. The dashed blue lines indicate the edges of the corresponding multi-phonon energy convolution (see text for more detail). The green lines are plots of the RIXS spectrum at the high symmetry points. }
\end{figure*}

\noindent\textit{Results and discussion} 
\label{sec:result} --- 
Figure~\ref{fig:1} presents RIXS spectra for an itinerant electron, Holstein-coupled to a dispersive optical phonon branch. Here we take $\omega_0 = 1$ as our unit of energy and set $\omega_1=0.02\omega_0$, $t = 5\omega_0$, $g=2\omega_0$, $U_Q = 20\omega_0$, and $\Gamma = 2\omega_0$ unless otherwise stated. The effective $e$-ph coupling $\lambda= M^2/4t \omega_0=0.2$ is thus rather weak. For a typical transition metal oxide we expect $\omega_0 \sim 100$~meV. Our value for $\Gamma$ is, therefore, halfway between values appropriate of the transition metal $L$-edge and oxygen $K$-edge~\cite{LeePRL2013, GeondzhianPRB2020, BraicovichPRR2020}. Our choice for $U_Q$ is smaller than the $U_Q\sim 4\textrm{--}6$~eV typically adopted in the literature \cite{OkadaPRB2001, KourtisPRB2012, LeePRL2013, TohyamaPRB2015, JohnstonNatComm2016}. This choice partially accounts for the interaction between the core hole and the lattice and extenuates the delocalization effects in the intermediate state. As discussed previously~\cite{BieniaszSciPost2021}, the core-hole-lattice coupling, which is neglected in Eq.~\eqref{eq:epcp}, can frustrate polaron formation in the valence band. At the lowest order, this effect \emph{reduces} the effective core hole potential, which we account for by reducing $U_Q$. 

For reference, Fig.~\ref{fig:1}(a) shows the RIXS spectrum for a dispersionless optical phonon ($\omega_1 = 0$). It shows the expected multi-phonon excitations located at multiples $n\omega_0$ of the phonon energy. The excitations have Lorentzian line shapes with a broadening set by $\eta=0.04\omega_0$ to mimic the instrument's resolution. The amplitude of the peaks decreases as the excitation number $n$ increases. (For more clarity, we scaled each overtone by the numerical factor indicated in red at the top of the plot.)  The momentum dependence of the intensity of the first phonon peak is due to the electron mobility in the intermediate state~\cite{BieniaszSciPost2021}. The intensity of the multi-phonon peaks is $\vect{q}$-dependent, but it is harder to discern on this scale. 

Figure~\ref{fig:1}(b) shows the RIXS spectrum when we introduce a phonon dispersion with a narrow bandwidth $\omega_{1}=0.02$. 
The single phonon peak continues to be a Lorentzian with broadening $\eta$, but its position now follows the phonon dispersion $\omega_{\vect{q}}$ indicated by the dashed blue line, as required by the conservation of momentum and energy \cite{AmentEPL2011, DevereauxPRX2016}. Its intensity again exhibits a significant momentum dependence due to the mobility of the electron in the intermediate state. This is further confirmed by the RIXS spectrum shown in Fig.~\ref{fig:1}(c) for a localized electron ($t = 0, \omega_1 = 0.02$). Indeed, here the single phonon peak tracks the phonon frequency $\omega_{\vect{q}}$ but has the same intensity at all $\vect{q}$. 

Much more important is the observation that now the higher-order peaks in Fig.~\ref{fig:1}(b) also show a strong momentum dependence both in their line shape and intensity. To understand it, consider first the two-phonon peak. Here, the total transferred momentum is distributed between the two phonons left behind after RIXS,  $\vect{q}=\vect{q}_1+\vect{q}_2$. The transferred energy must then equal the two phonons' energy 
$ \omega_{\vect{q}-\vect{q}_2}+\omega_{\vect{q}_2}
= 2\omega_{0}+ 4\omega_{1}\sum_{\delta=x,y}\cos\frac{q_{\delta}}{2}\cos(q_{2_\delta}-\tfrac{q_{\delta}}{2})$. There is no broadening at the $M$-point [$=(\pi,\pi)$] (apart from the extrinsic broadening $\eta$); however, for any other $\vect{q}$ the two-phonon peak has an intrinsic broadening $8\omega_1 (\cos\frac{q_x}{2} +\cos\frac{q_y}{2})$, marked by the blue dashed lines, due to the convolution over all $\vect{q}_2$ values. 

We can explain the broadening of the higher multi-phonon peaks in a similar manner; it results from the convolution over the $n$ phonon energies with total momentum $\vect{q} = \sum_{i=1}^n \vect{q}_i$. The expected outermost energies allowed by this constraint are shown by the dashed blue lines and indeed mark the regions with finite RIXS intensity. The higher-order peaks thus exhibit an ever-growing broadening. For example, the four-phonon feature is approximately twice as wide as the two-phonon one. Fig.~\ref{fig:1}(c) shows the same broadening for the localized electron, confirming that this feature is due solely to the phonon dispersion. These findings naturally explain why many experiments have resolved increasing line widths for the multi-phonon excitations \cite{LeePRL2013, JohnstonNatComm2016}. 

Another interesting observation is that the shape of the two- and three-phonon peaks is highly nontrivial and does not follow a Lorentzian or Gaussian lineshape, as is often assumed. Furthermore, the three phonon peak is skewed, producing asymmetric peaks around the $\Gamma$- and $M$-points. The bond-stretching ``breathing'' phonon modes in transition metal oxides often have bandwidths comparable to our model \cite{Pintschovius2005} while Cu $L$-edge RIXS experiments can access momentum transfers approaching the $X$ point. Therefore, copper oxide materials could serve as a platform for experimentally confirming these effects, provided the coupling is strong enough to generate multi-phonon excitations and depending on the instrumental resolution and the actual self-energy broadening of the valence electron. Nonetheless, it seems to be worth investigating. For example, it might be worth examining how incoherent, extremely correlated Fermi liquid \cite{ShastryPRL2011, MaiPRB2018} or non-Fermi liquid \cite{ReberNatPhys2012} behavior would manifest here. 

The higher phonon excitations eventually revert to a Gaussian line shape, as evident in the four-phonon line, owing to the central limit theorem (the crossover from unusual to Gaussian line shapes is controlled by the strength of the $e$-ph coupling). In contrast to the case of a dispersionless phonon (panel a), we also see a stronger momentum dependence of the weight of the higher-phonon peaks. Comparison with panel (c) reveals that its details depend on the phonon and valence band's bandwidth. We expect that this intensity and the specific line shapes will be further affected by a momentum-dependence of the $e$-ph coupling, but the study of this issue is deferred to future work.

\begin{figure}[t!]
\includegraphics[width=\linewidth]{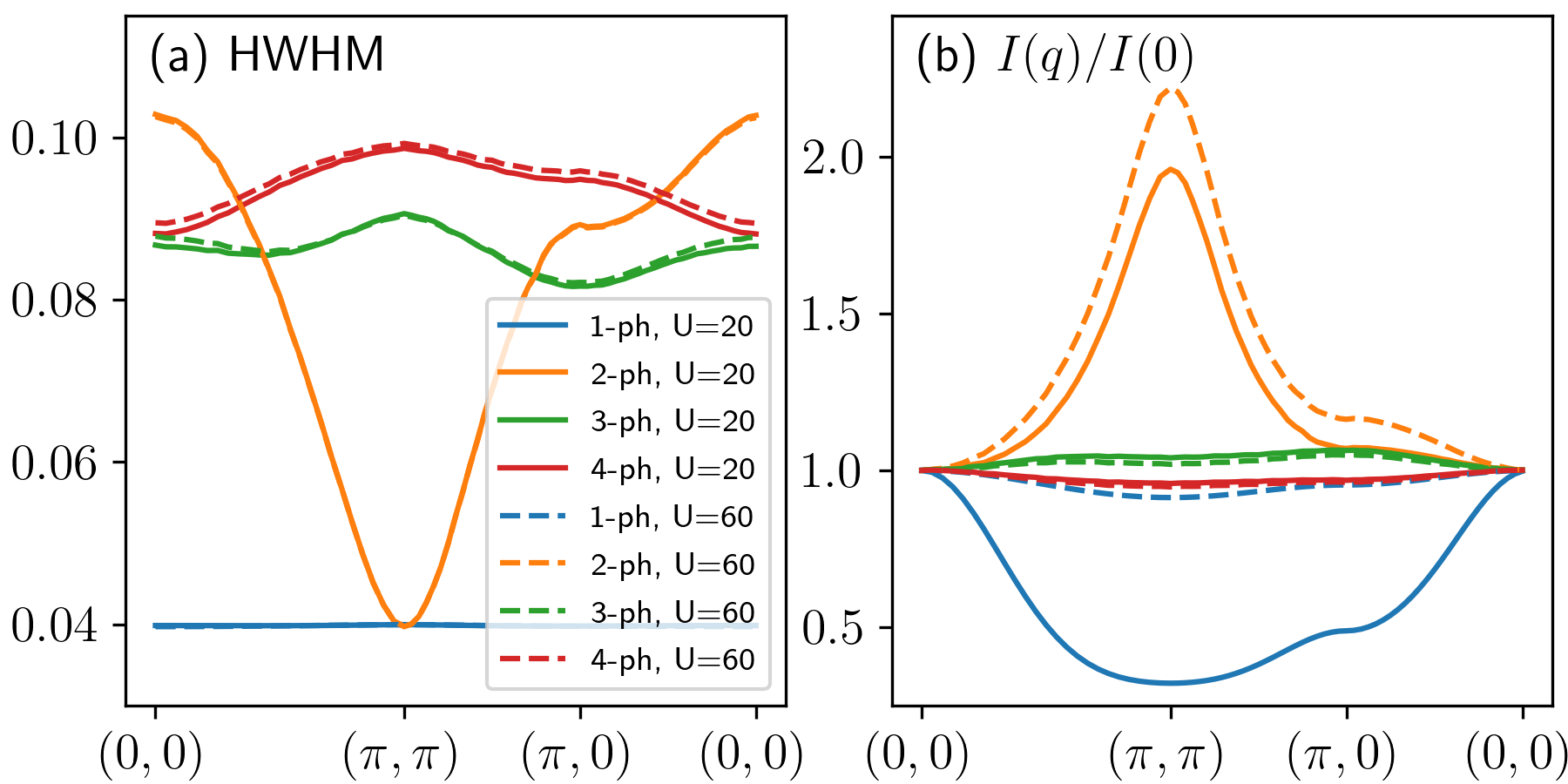}
\caption{\label{fig:2} 
An analysis of the phonon excitations for $\omega_0 = 1$, $t = 5\omega_0$, $\omega_1 = 0.02\omega_0$, $\Gamma = 2\omega_0$, and $g = 2\omega_0$. (a) The width of the phonon excitations, defined as the empirically determined half width half maximum (HWHM) of the peak, as a function of momentum. (b) The momentum dependence of the maximum phonon excitation intensity, normalized to their peak intensity at $(0,0)$. In all panels, results are shown for the first four phonon excitations and for $U_Q = 20\omega_0$ (solid lines) and $U_Q = 60\omega_0$ (dashed lines).}
\end{figure}

Figure \ref{fig:1} clearly illustrates that both electron mobility and phonon dispersion, and their interplay, produce phonon excitations with nontrivial momentum-dependence in the RIXS spectra. We further quantify these results in Fig.~\ref{fig:2} for different values of $U_Q$. Figures ~\ref{fig:2}(a) and ~\ref{fig:2}(b) plot the line width of the phonon peaks and their intensity as a function of momentum, as obtained from numerical fitting of a Lorentzian lineshape. Here, the peak intensity is determined from the peak maximum. One could use the integrated area instead, which would show similar trends but with quantitative differences (not shown). The width is determined empirically from the half-width at half maximum (HWHM). For $U_Q = 20\omega_0$, the intensity of the first and second phonon excitations varies significantly. For example, the first phonon excitation drops in intensity by more than half when tracking from $(0,0)$ to $(\pi,\pi)$, while the intensity of the second phonon excitation grows by a factor of two. As discussed, the width of the first phonon peak is fixed to our input resolution ($\eta=0.04\omega_0$), while the width of the second phonon peak varies by more than 100\% following the trends noted previously. In comparison, the momentum dependence of the third and fourth phonon excitations is weaker but remains significant.

\begin{figure}
    \centering
    \includegraphics[width=\linewidth]{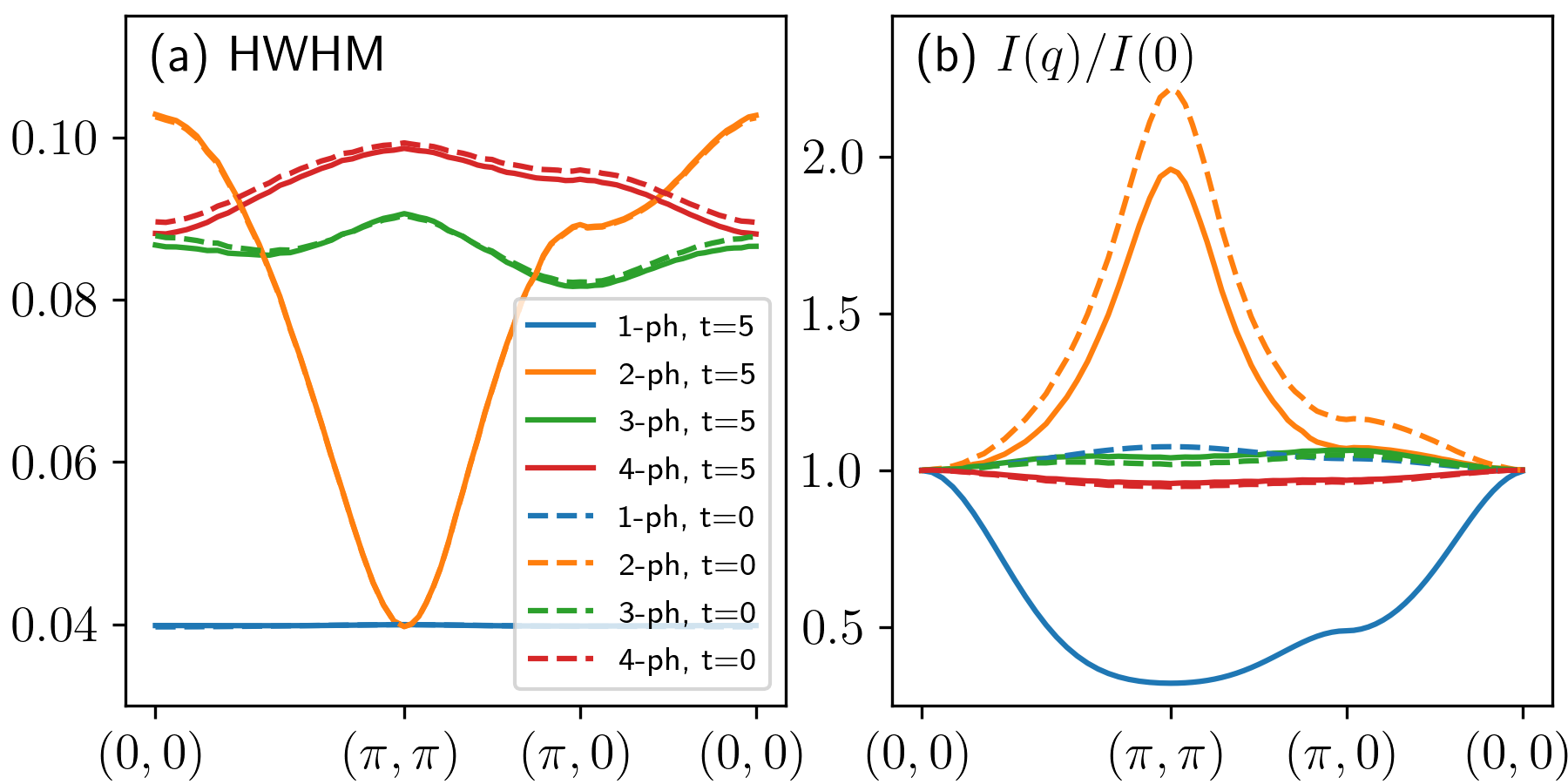}
    \caption{\label{fig:3} 
    An analysis of the phonon excitations for localized ($t = 0$, dashed lines) and delocalized ($t = 5\omega_0$, solid lines) electrons, similar to Fig.~\ref{fig:2}. Other parameters are 
    $\omega_0 = 1$, $\omega_1 = 0.02\omega_0$, $U_Q = 20\omega_0$, and $g = 2\omega_0$. 
    }
\end{figure}

We also examine a larger core-hole potential $U_Q = 60\omega_0$. This value effectively localizes the excited valence electron at the core-hole site in the intermediate state and reduces the momentum dependence of the phonon peaks (similar to the $t=0$ results). The only exception is the two-phonon peak, which still varies rapidly as a function of~$\vect{q}$. This result suggests that the largest contribution to the momentum dependence of two-phonon excitation arises from the phonon dispersion rather than the electron mobility. To confirm this, Fig.~\ref{fig:3} compares the results of the 
same analysis, this time for systems with $t = 0$ and $t=5\omega_0$, and $U_Q = 20\omega_0$.  The strong similarity between the results for the localized electron ($t = 0$) and those obtained for a mobile electron with $U_Q = 60\omega_0$, indicate that this larger potential is indeed strong enough to localize the electron in the intermediate state.\\

\noindent\textit{Summary and Conclusions} --- Our results demonstrate that electron mobility and phonon dispersion produce momentum-dependent phonon excitations in the RIXS spectra. Crucially, this dependence emerges even for models with \emph{momentum independent} $e$-ph interactions and would significantly impact estimates for the strength of the coupling if one does not account for it.  We also found that the phonon dispersion produces a nontrivial broadening of the multi-phonon excitations. This effect may account for the increasing widths of the phonon excitations often observed in experiments \cite{LeePRL2013, JohnstonNatComm2016}. 

Our results have important implications for analyzing RIXS data on systems with dispersive phonon modes. For example, the single-site model predicts that the strength of the $e$-ph coupling can be directly extracted from the intensity ratios of successive phonon excitations with $g_{\vect q}/\Gamma = I_{n+1}({\vect q})/I_n({\vect q})$ \cite{AmentEPL2011}. In our model, $g$ is independent of $\vect{q}$ yet the resulting $I_n(\vect{q})$ are not, showing that this simple mapping does not hold for dispersive systems. Importantly, this conclusion holds in the limit of strong electron localization (i.e., strong, attractive core-hole potentials) if the relevant phonon branch has a sizable bandwidth.  

Our results are for a band insulator, where the core electron is excited into an empty band in the intermediate state of the scattering process. We suggest that such systems, along with other dilute materials, could be used to develop a controlled theory of $e$-ph coupling in RIXS experiments. In the future, it is highly desirable to explore these effects in cases where the band is partially filled and/or where correlation effects cannot be neglected.\\

\begin{acknowledgments}
\noindent\textit{Acknowledgments} --- K.~B. and M.~B. acknowledge support from the UBC Stewart Blusson Quantum Matter Institute (SBQMI) and the Natural Sciences and Engineering Research Council of Canada (NSERC). S.~J. is supported by the National Science Foundation under Grant No. DMR-1842056.
\end{acknowledgments}

\bibliographystyle{apsrev4-1}
\bibliography{rixs}

\newpage

\section{Supplementary material: Derivation of the variational method for the inhomogeneous case}

Here we provide the technical details of the inhomogeneous partial fractions solution to Eq.~(1) of the main text, a generalization of the method presented in \cite{Ber10,Ebr12} that is necessary to obtain the phonon momentum dependence. Throughout, we partition the Hamiltonian $\mathcal{H} = \mathcal{H}_\mathrm{e}+\mathcal{H}_\mathrm{ph}+\mathcal{H}_{\textrm{$e$-ph}}+\mathcal{H}_\mathrm{ch}$, where 
\begin{eqnarray*}\nonumber
    \mathcal{H}_{e}&=& -t\sum_{\mean{ij}}[d_{i}^{\dag}d_{j}^{\pdag}+\mathrm{H.c.}]\\
    \mathcal{H}_\mathrm{ph}&=&\sum_{\vect{q}} \omega^{\phantom\dagger}_{\vect{q}} b_{\vect{q}}^{\dag}b_{\vect{q}}^{\pdag}\\
    \mathcal{H}_{\textrm{$e$-ph}}&=&\frac{g}{\sqrt{N}} \sum_{i} e^{-\mathrm{i}\vect{k}\cdot\vect{R}_{i}} d_{i}^{\dag}d_{i}^{\pdag}(b_{\vect{k}}^{\dag}+b_{-\vect{k}}^{\pdag}),~\mathrm{and}\\
    \mathcal{H}_\mathrm{ch}&=&\epsilon_\mathrm{ch}\sum_{i}p_i^\dag p^{\pdag}_i -U_Q\sum_{i}d_{i}^{\dag}d_{i}^{\pdag}\left(1-p_i^\dag p^{\pdag}_i\right), 
\end{eqnarray*}
and all of the operators and parameters are defined as in the main text.

Our starting point is the generalized propagator
\begin{equation}
  \label{eq:spec}
  \mathcal{F}_{fg}(\vect{q},z) = \frac{1}{\sqrt{N}}\sum_{i}e^{\mathrm{i}\vect{q}\cdot{\vect R}_{i}} \bra{f}p_{i}^{\dag}d_{i}^{\pdag} \mathcal{G}(z) d_{i}^{\dag}p_{i}^{\pdag}\ket{g},
\end{equation}
which is the main physical quantity for calculating the scattering amplitude for the RIXS cross-section $F_{fg}(\vect{q},z)$. Because we consider RIXS in an insulator,  $\ket{g}$ is a state with all core states filled and the valence band empty. For simplicity, we choose $ \ket{g} \equiv \ket{0}$ to be the phonon vacuum. X-ray absorption excites an electron from the $p$-core level into the $d$-valence band: $d_{i}^{\dag}p^{\phantom\dagger}_{i}\ket{g}$. At the end of the process, the valence electron decays into the core level and the system is left in a multi-phonon state $\ket{f}$.

Let us first introduce the following definition for the generalized Green's function to simplify the notation:
\begin{multline}
  \mathcal{F}_{fn}^{ij}(z,\Delta)=
  \frac{1}{N^{\frac{n+1}{2}}}\sum_{\set{\vect{k}}^{n}}
  e^{\mathrm{i}\vect{q}\cdot\vect{R}_{i}}e^{-\mathrm{i}\ssum_{l}\vect{k}_{l}\cdot\vect{R}_{j}}\\
   \quad\quad\times\bra{f} d_{i}^{\pdag}\mathcal{G}^{\pdag}_i(z)d_{j}^{\dag}b_{\vect{k}_{1}}^{\dag}\ldots b_{\vect{k}_{n}}^{\dag}\ket{0},\label{eq:Fn}
\end{multline}
where $\Delta=\vect{R}_{j}-\vect{R}_{i}$ 
denotes a 2D vector pointing from the core hole site to any other site, $N\to\infty$ is the total number of sites  and  $\set{\vect{k}}^{n}\equiv\set{\vect{k}_{1},\ldots,\vect{k}_{n}}$ for brevity. The presence of the core-hole at site $i$ results in an on-site attraction $-U_Q d^{\dag}_id^{\phantom\dagger}_i$, whose presence is marked by labeling the resolvent $\mathcal{G}(z) \rightarrow  \mathcal{G}_i(z)$. This removes the need to specify explicitly that the core-hole is at site $i$ throughout the remaining calculations.
The spectral amplitude of Eq.~\eqref{eq:spec} is identified with $\mathcal{F}_{f0}^{ii}(z)$.

Applying the Dyson identity for $\mathcal{V}=\mathcal{H}_{e-\textrm{ph}}$ and $\mathcal{H}_{0}=\mathcal{H}_{e}+\mathcal{H}_{\textrm{ph}}+\mathcal{H}_\mathrm{ch}$, we arrive at the zero order EOM:
\begin{equation}
  \label{eq:eom1}
  \mathcal{F}_{f0}^{ii}(z) = \sqrt{N} G_{0}^{ii}(z)\delta_{\vect{q}0}\delta_{f0} + M\sum_{l}\mathcal{F}_{f1}^{il}(z,\Delta)G_{0}^{li}(z),
\end{equation}
where $G_0^{li}(z)$ is the bare propagator for the $d$-electron in the presence of the core-hole attraction $-U_Q d_i^\dag d^\pdag_i$. (Details on how to compute this propagator can be found in the appendix of Ref.~\citenum{BieniaszSciPost2021}.) 
The first term in Eq.~\eqref{eq:eom1} is responsible for the elastic scattering and produces the zero phonon peak in the RIXS spectrum. (As is customary, we omit it in our numerical results as it tends to obfuscate the lattice excitations.) The second term contains the one phonon Green's function, which needs to be expanded using the Dyson equation. This EOM is exact. 

The higher order propagators' EOM  can be obtained similarly, and then simplified to:
\begin{multline}
  \label{eq:eomn}
  \mathcal{F}_{fn}^{il}(z,\Delta) \approx 
  g\Big[\mathcal{F}_{fn+1}^{il}(z,\Delta)+n\mathcal{F}_{fn-1}^{il}(z,\Delta)\Big]\mean{G_{0}^{ll}(z)}_{n}\\
  +n! B_{fn}^{il}(z)
\end{multline}
where the inhomogeneous term is
\begin{equation}
  \label{eq:B}
B_{fn}^{il}(z) = \frac{1}{N^{\frac{n-1}{2}}}\sum_{\set{\vect{k}}^{n}}e^{-\mathrm{i}\vect{q}\cdot\Delta}G_{0}^{il}(z-{\ssum}_{m}\omega_{\vect{k}_{m}})\delta_{\vect{q},\ssum_{m}\vect{k}_{m}}\delta_{f,\set{\vect{k}}^{n}}.
\end{equation}
To obtain this result, we have already employed the MA approximation, namely the $\mathcal{F}_{fn\pm1}$ functions have been decoupled from the free propagators $G_{0}(z)$ by $n$-fold averaging the latter over the first Brillouin zone~\cite{BerciuPRL2006}
\begin{equation}
  \label{eq:G0n}
  \mean{G_{0}^{ij}(z)}_{n} = \frac{1}{N^{n}}\sum_{\set{\vect{k}}^{n}}G_{0}^{ij}(z-\ssum_{l}\omega_{\vect{k}_{l}}).
\end{equation}
Mathematically, this is equivalent to only keeping configurations where the $n$ phonons are all at the same site in real space. Another thing to notice is the presence of the $\delta_{\vect{q},\ssum_{m}\vect{k}_{m}}$ factor, a simplification which is possible due to the fact that MA employs infinite lattice Green's functions. Thus, the propagator $G_{0}^{ii}(z)$ is independent of the location of the core hole site~$i$ and the sum over the lattice in Eq.~\eqref{eq:spec} reduces to a simple conservation of momentum.

Equation (\ref{eq:G0n}) reflects the main change, within the MA approximation, coming from using dispersive phonons. For the case of non-dispersive (Einstein) phonons with $\omega_{\vect{q}}\equiv \omega_0$ discussed in Ref. \onlinecite{BieniaszSciPost2021}, the average over the $n$ phonons' momenta is trivial:  $\mean{G_{0}^{ij}(z)}_{n} \rightarrow G_{0}^{ij}(z-n \omega_0)$. In the presence of dispersive phonons, the momentum averages of Eq. (\ref{eq:G0n}) must be calculated numerically instead. Apart from this difference, the formal solution follows that discussed in Ref. \onlinecite{BieniaszSciPost2021}. For completeness, we briefly review it below.

Next, we solve this variationally simplified system of coupled EOMs by means of the continued fraction method. For a final state $|f\rangle$ with $n_{\!f}$ phonons, the continued fraction is calculated in the usual manner \cite{Ber10} by positing that
\begin{equation}
  \label{eq:conf}
  \mathcal{F}_{fn}^{il}(z)=A_{n}^{l}(z)\mathcal{F}_{f,n-1}^{il}(z)
\end{equation}
for $n>n_{\!f}$. This equation has the standard solution
\begin{equation}
  \label{eq:An}
  A_{n}^{l}(z) = \frac{ng\mean{G_{0}^{ll}(z)}_{n}}{1-g\mean{G_{0}^{ll}(z)}_{n}A_{n+1}^{l}(z)},
\end{equation}
which can be calculated recursively with the additional physical constraint $\lim_{n\to\infty}A_{n}^{l}(z)=0$.

For $n=n_{\!f}$, the EOM contains the inhomogeneous term $n!B_{fn_{\!f}}^{il}(z)$, which necessitates that the continued fraction takes the form
\begin{equation}
  \label{eq:confr}
  \mathcal{F}_{fn}^{il}(z)=\bigg[\mathcal{F}_{f,n-1}^{il}(z) + \frac{(n-1)!B_{fn_{\!f}}^{il}(z)}{g\mean{G_{0}^{ll}(z)}_{n_{\!f}}}\bigg]A_{n}^{l}(z),
\end{equation}
and similarly thereafter until the $\mathcal{F}_{f0}^{il}(z)$ function is reached. Ultimately, the required zero order function $\mathcal{F}_{f0}^{ii}(z)$ can be derived from the self-consistent equation
\begin{equation}
  \label{eq:F0}
  \mathcal{F}_{f0}^{ik}(z) = \sum_{l}\Big[
  g\mathcal{F}_{f0}^{il}(z)A_{1}^{l}(z) + \mathcal{B}_{fn_{\!f}}^{il}(z)
  \Big] G_{0}^{lk}(z),
\end{equation}
where the free coefficients
\begin{align}
  \label{eq:pi}
  \mathcal{B}_{fn_{\!f}}^{il}(z) &= \frac{B_{fn_{\!f}}^{il}(z)}{\mean{G_{0}^{ll}(z)}_{n_{\!f}}}\mathcal{A}_{n_{\!f}}^{l}(z),~\mathrm{and}\\
  \mathcal{A}_{n_{\!f}}^{l}(z) &= \prod_{m=1}^{n_{\!f}}\!A_{m}^{l}(z)
\end{align}
result from the chain of EOMs linking the $\mathcal{F}_{fn_{\!f}}^{il}(z)$ and $\mathcal{F}_{f0}^{il}(z)$ functions. Here, the $l$ summation is over all the sites in the system where the polaron cloud can appear. This fact turns the implicit Eq. (10) into another infinite
system of coupled equations (when $N\rightarrow\infty$); however, this can be recast as a different sum over all sites where the contributions decay very fast to zero as the distance $|l-i|$ increases. Effectively, this latter sum is then replaced with a finite sum over sites within a Manhattan distance $p$ of the core-site, which can be solved efficiently. (The cutoff value $p$ is increased until convergence is achieved, usually $p=2$ suffices.) This recasting of the original sum over $l$ is achieved by renormalizing the energy $z$ by the self-energy of a free Holstein polaron, which accounts for the contribution from sites $|l-i|> p$. This step is identical to that used in our previous work on the Einstein phonon RIXS theory \cite{BieniaszSciPost2021}, and we do not repeat it here. By comparing with that paper, we see that the dispersive phonon effects are contained in the $\mathcal{B}_{fn_{\!f}}^{il}(z)$ factor, which for Einstein phonons reduces to $\delta^{il}$.

The explicit solution of the RIXS spectral function can be cast in matrix form as
\begin{align*}
  \label{eq:sol}
  \mathcal{F}_{f0}^{ii} &= \mathcal{B}_{fn_{\!f}}^{il}(z)
  T^{lj}(z) G_{0}^{ji}(z),\\
  T^{lj}(z) &= \delta^{lj} + gG_{0}^{lk}(z)A_{1}^{k}(z) \big[\delta^{kj}-gG_{0}^{kj}(z)A_{1}^{j}(z)\big]^{-1},
\end{align*}
where the $T^{lj}(z)$ transformation is the solution to the linear system in Eq.~\eqref{eq:F0}.

Finally, we must use the above solution to calculate the RIXS cross-section $I(\omega,\vect{q},z)$. Recall that the free coefficient $B_{fn_{\!f}}$ depends on the $\delta_{f,\set{\vect{k}}^{n}}$ factors, which cannot be integrated over explicitly. Rather, we can find the $n$-phonon contribution to the cross-section
\begin{widetext}
\begin{equation}
  \label{eq:rixs}
   I_{n}(\omega,\vect{q},z) = \bigg[\frac{G_{0}^{ij'}(z)T^{j'l'}(z)\mathcal{A}_{n}^{l'}(z)}{\mean{G_{0}^{l'l'}(z)}_{n}}\bigg]^{*}
  \mathcal{I}_{n}^{l'l}(\omega,\vect{q},z)
  \frac{\mathcal{A}_{n}^{l}(z)T^{lj}(z)G_{0}^{ji}(z)}{\mean{G_{0}^{ll}(z)}_{n}},
\end{equation}
and the integral over all the final states $f$ corresponding to the $n$-phonon RIXS contribution can be simplified to
\begin{equation}
  \label{eq:nph}
  \mathcal{I}_{n}^{l'l}(\omega,\vect{q},z) =
%\sum_{f} \frac{{[B_{fn}^{l'i}]}^{*}B_{fn}^{il}}{\omega+i\eta-\omega_{f}} =
  \frac{1}{N^{n-1}}\sum_{\set{\vect{k}}^{n}} \frac{ {[G_{0}^{l'i}(z-\ssum_{j}\omega_{\vect{k}_{j}})]}^{*}G_{0}^{il}(z-\ssum_{j}\omega_{\vect{k}_{j}}) e^{\mathrm{i}\vect{q}\cdot(\vect{R}_{l}-\vect{R}_{l'})}\delta_{\vect{q},\ssum_{j}\vect{k}_{j}}}{\omega+\mathrm{i}\eta-\ssum_{j}\omega_{\vect{k}_{j}}}.
\end{equation}
\end{widetext}
The $1/N^{(n-1)}$ prefactor is indeed the proper normalization since Eq.~\eqref{eq:nph} involves a $(n-1)$-fold integration after the momentum conservation is explicitly evaluated. This integration is the main numerical challenge of this method. We can efficiently approximate it owing to the assumption of weakly dispersive optical phonons, leading to a numerical integration over a very narrow density of states.
Since the energy shifts $\omega_{\vect{k}}$ over a moderately dense mesh of $k$-values will be strongly localized and many of them will in fact be identical, approximating the $n$-phonon density of states becomes a relatively simple numerical task.

Finally, the RIXS cross-section is the sum of all the $n$-phonon contributions, up to the desired level (in our case, up to 4 phonons), and taken at a particular resonance energy $z_{r}$
\begin{equation}
  \label{eq:csec}
  I(\omega,\vect{q}) = -\frac{1}{\pi}\Im\sum_{n}I_{n}(\omega,\vect{q},z_{r}).
\end{equation}

\end{document}